\documentclass[twocolumn,twoside,slac]{revtex4}
\usepackage{graphicx}
\usepackage{fancyhdr}
\begin{document}

\title{Architecture of the ATLAS High Level Trigger Event Selection Software}

%

\author{S. Armstrong, K.A. Assamagan, J.T. Baines, C.P. Bee, M. Biglietti, A. Bogaerts, 
V. Boisvert, M. Bosman, S. Brandt, B. Caron, P. Casado, G. Cataldi, D. Cavalli, M. Cervetto, 
G. Comune, A. Corso-Radu, A. Di Mattia, M. Diaz Gomez, A. dos Anjos, J. Drohan, N. Ellis, M. Elsing, B. Epp, F. Etienne, S. Falciano, A. Farilla, S. George, V. Ghete, 
S. Gonzalez, M. Grothe, A. Kaczmarska, K. Karr, A. Khomich, N. Konstantinidis, W. Krasny, W. Li, A. Lowe, L. Luminari, H. Ma, C. Meessen, A.G. Mello, G. Merino, P. Morettini, E. Moyse, A. Nairz, A. Negri, N. Nikitin, A. Nisati, C. Padilla, F. Parodi, 
V. Perez-Reale, J.L. Pinfold, P. Pinto, G. Polesello, Z. Qian, S. Rajagopalan, S. Resconi, S. Rosati, D.A. Scannicchio, C. Schiavi, T. Schoerner-Sadenius, E. Segura, T. Shears, 
S. Sivoklokov, M. Smizanska, R. Soluk, C. Stanescu, G. Stavropoulos, S. Tapprogge, F. Touchard, 
V. Vercesi, A. Watson, T. Wengler, P. Werner, S. Wheeler, F.J. Wickens, 
W. Wiedenmann, M. Wielers, H. Zobernig \\
\vspace*{0.2cm} 
The ATLAS High Level Trigger Group\footnote{http://atlas.web.cern.ch/Atlas/GROUPS/DAQTRIG/ \\
HLT/AUTHORLISTS/chep2003.pdf} \\
\vspace*{0.2cm}
Presented by Monika Grothe}
\affiliation{European Organization for Nuclear Research, CERN, 
1211 Geneva 23, Switzerland}

\begin{abstract}
The ATLAS High Level Trigger (HLT) consists of two selection steps: the second level trigger and the event filter. Both will be implemented in
software, running on mostly commodity hardware. Both levels have a coherent approach to event selection, so a common core software framework has
been designed to maximize this coherency, while allowing sufficient flexibility to meet the different interfaces and requirements of the two different
levels. The approach is extended further to allow the software to run in an off-line simulation and reconstruction environment for the purposes of
development. This paper describes the architecture and high level design of the software.
\end{abstract}

\maketitle

\thispagestyle{fancy}


\section{INTRODUCTION}

The Large Hadron Collider LHC, currently under construction at CERN and 
scheduled to start data-taking in 2007, will collide protons on protons at a 
center-of-mass energy of 14 TeV. At the design luminosity of 
$10^{34} {\rm cm^{-2} s^{-1}}$, each bunch crossing, occurring at 25~ns 
intervals,
will result in 23 collisions on average.

The trigger of the ATLAS experiment, one of the multi-purpose detectors at 
the LHC, must be able to reduce the interaction rate of $10^9$~Hz
to below the maximum rate that can be processed by the off-line computing 
facilities, $\mathcal{O}(10^2)$~Hz. In addition, the ATLAS trigger must be
able to handle the full data volume of a detector of the complexity and size of
ATLAS, with $\mathcal{O}(10^8)$ read-out channels. It is under these constraints
that the ATLAS trigger must retain the capability of
identifying previously undetected and rare physics processes. A Standard Model
Higgs particle with a mass of 120~MeV, decaying into two photons, is for 
example
expected to occur at a rate of $10^{-13}$ of the interaction rate, the 
proverbial pin in the haystack.

Three distinct trigger steps are foreseen. While the first step 
(Level-1 trigger) is implemented as a hardware trigger, the second and third 
steps, Level-2 trigger and Event Filter, are software triggers and are 
usually referred to as the ATLAS High Level Trigger (HLT). 

This article describes the current status of the high-level design and 
the implementation of 
the HLT event selection software. The validation of the software is on-going
and aims at the imminent goal of the HLT Technical Design Report (TDR), due 
in summer 2003, as well as at preparing for the first stage of ATLAS 
commissioning.

After a short overview of the ATLAS trigger, the
general strategy of the HLT event selection software and its main software
components are described. Subject of this article is the part of the code that 
provides the infrastructure in which to run the HLT selection algorithms.
Specific HLT event selection algorithms are not discussed.
Using the same software environment to develop and maintain 
(off-line) on the one hand and on the other hand to deploy (on-line) the HLT 
selection software is a central HLT design aim. 
This article also discusses our experience with appropriating off-line code for
on-line use.

\section{THE ATLAS TRIGGER}

\begin{figure}[tb]
\includegraphics[width=65mm]{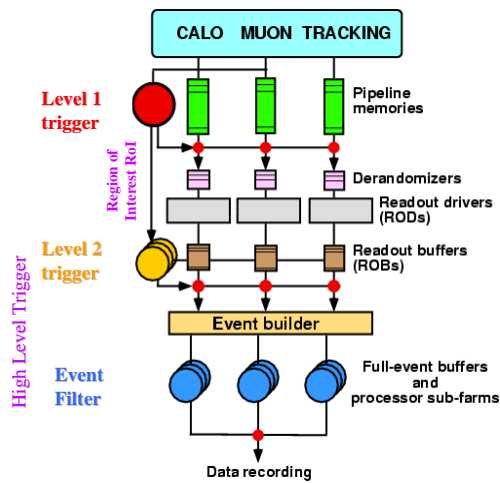}
\caption{Overview of the ATLAS trigger} 
\label{proc0}
\end{figure}

Data at the LHC are produced with the bunch crossing rate of 40~MHz.
The ATLAS trigger (see Fig.~\ref{proc0}) has the task to reduce this rate to 
an output rate of $\mathcal{O} (200)$~Hz after the Event Filter. The 
Level-1 trigger~\cite{LVL1TDR} 
reduces the initial 40~MHz to less than 75~kHz in less than 2.5~$\mu$s, the 
maximum output rate and latency the
trigger hardware can tolerate. In the HLT~\cite{TP}, where the boundary 
between 
the two trigger steps is purposefully kept flexible, the Level-2 trigger will
reduce the rate to $\mathcal{O}(2)$~kHz and the Event Filter further to 
$\mathcal{O}(200)$~Hz. The available average latency of the two steps is
substantially different, with $\sim$10~ms for the Level-2 trigger and 
$\sim$1~s for the Event Filter.

Central to the ATLAS trigger design is the Region-of-Interest (RoI) concept.
The Level-1 trigger looks for regions of potentially interesting activity in 
the Calorimeters and the Muon Spectrometer that may correspond to candidates 
for high $p_T$ objects (electromagnetic, muon, tau/hadronic and jet clusters).
The Level-2 trigger selection uses the RoI information (its type, position and 
the $p_T$ of the highest trigger threshold passed) of the Level-1 
trigger as seeds for its processing. This strategy keeps the amount of raw 
data to be passed to the Level-2 trigger for processing at only a few 
per cent of the full event information.  

While the raw data of the ATLAS subdetectors are held in readout buffers (ROB),
the Level-2 trigger 
processes RoI information and raw data on a Level-2 Linux PC farm.
On each node of the Level-2 farm, a Processing Application executes the 
Level-2 event selection software. Specialized, speed-optimized Level-2 
trigger algorithms 
carry out the feature extraction in the region-of-interest indicated by the
Level-1 trigger. While the Level-1 trigger does not combine the information
from different subdetectors, the Level-2 trigger covers all subdetectors
sequentially and gains additional information from combining their data.

Once an event has been accepted by the Level-2 trigger, the raw data of the 
full event are passed to the Event Builder, which in turn passes on the 
fully-built 
event to the Event Filter Linux PC farm. On the farm nodes, independent 
Processing Applications execute off-line-type selection algorithms that have 
access to the full event data, including the latest calibration and alignment 
information.

Though the type of selection algorithms used by the Level-2 trigger 
and the Event Filter are different, the software infrastructure that 
the algorithms use is kept as similar as possible between the two trigger 
levels. 
In order to achieve a high level of flexibility, it is desirable to retain to
some degree the possibility to deploy an HLT algorithm either in the Level-2
trigger or in the Event Filter. This article describes the common software
infrastructure that is used throughout the HLT. Given the much more stringent
performance constraints of the Level-2 trigger, this approach meets limits
that will be discussed as well.

\section{THE HIGH LEVEL TRIGGER EVENT SELECTION SOFTWARE}

The HLT event selection software has four main components:
the HLT algorithms that perform event reconstruction and feature extraction, 
the HLT Steering that calls a certain subset of the available HLT algorithms 
in a certain sequence depending on the types of RoI received from the Level-1
trigger, the HLT Data Manager that gives access to the information contained 
in the raw data, and the HLT raw Event Data Model that specifies the object
representation of the raw data to be used by the HLT algorithms. This section
describes how the four components work together and discusses the design and 
current implementation of the three components, HLT Steering, HLT Data Manager,
HLT Event Data Model, that provide the infrastructure for the HLT algorithms
both in the Level-2 trigger and in the Event Filter.

\subsection{Working principle}

\begin{figure*}[tb]
\includegraphics[width=135mm]{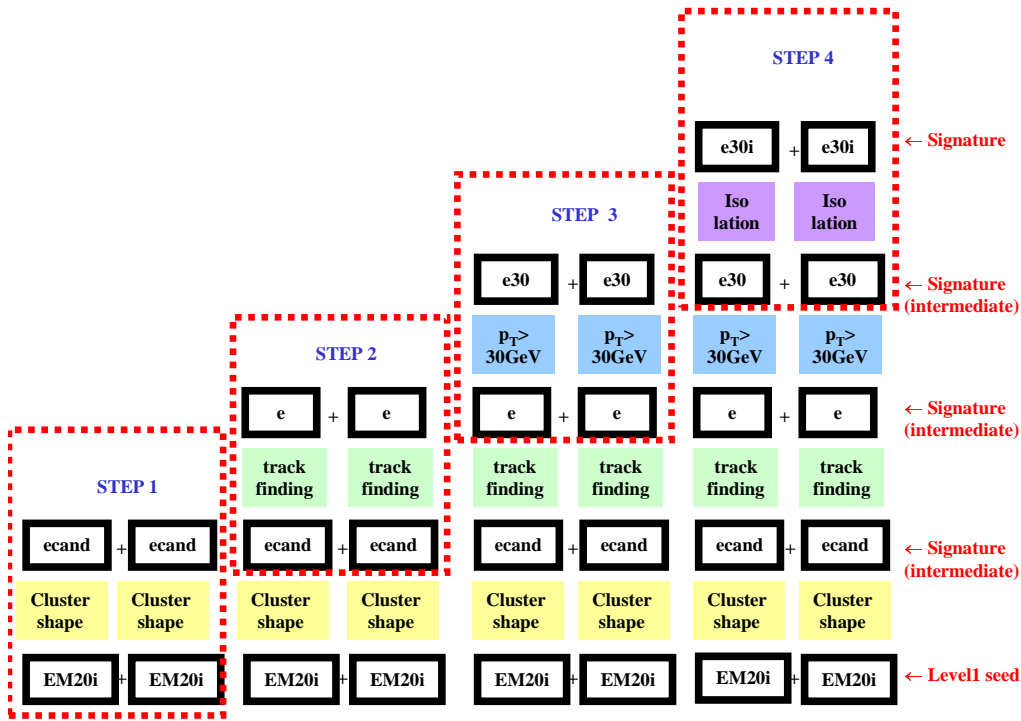}
\caption{Example illustrating the working principle of the HLT selection 
software. The different steps follow each other in time. Each step is shown 
with the complete processing chain preceding it. Hence for step 4, the full 
chain of processing steps is shown, with time increasing from bottom to top.} 
\label{proc1}
\end{figure*}

The following example illustrates the working principle of the HLT event 
selection software: The Level-1 trigger finds two isolated electromagnetic 
clusters with $p_T > 20$~GeV each. This is a possible signature for the
decay $Z \rightarrow e^+ e^-$ for which the ATLAS trigger accepts an event, 
if both electrons are isolated and have a minimum $p_T$ of 30~GeV.

The HLT validates this hypothesis in a step-by-step process. 
Intermediate signatures are produced as result of the algorithmic processing 
of a given step, and each intermediate signature is examined in order to be 
able to reject the hypothesis at the earliest possible moment. This procedure
is managed by the HLT Steering package.

The validation procedure starts from the Level-1 RoI information as seed, as 
shown in Fig.~\ref{proc1}.
In this example, there are two isolated electromagnetic RoIs with $p_T>20$~GeV
(``EM20i'' in Fig.~\ref{proc1}).
The HLT Steering calls an HLT algorithm (``Cluster Shape'') to determine the cluster shape of the 
first seed. The cluster shape is found consistent with the 
electron
hypothesis and the result is an electron candidate (``ecand''). 
The HLT Steering 
proceeds in the same way with the second seed and obtains a second 
electron candidate. These two electron candidates are considered 
as an intermediate signature (``ecand + ecand''). Since the event after this 
first step is still
compatible with the $Z \rightarrow e^+ e^-$ signature, the HLT Steering
starts the second step of the validation procedure. In this second step,
the two electron candidates provide the seeds for the algorithmic
processing. The HLT Steering calls a track finding HLT algorithm 
(``track finding'') for each
of the two electron candidates. If for each of the electron candidates a track
pointing to them is found, this second step results in two electrons that
constitute another intermediate signature (``e + e''). 
Using, in the manner described above, the output of one step of the validation 
procedure as input to the next,
the HLT Steering calls the appropriate HLT algorithms in the appropriate 
sequence to verify that the two electrons each have $p_T>30$~GeV and are both
isolated. When finally, after the fourth step, the signature ``e30i + e30i'' 
has been reached, the ATLAS trigger accepts the 
event as a candidate for the decay $Z \rightarrow e^+ e^-$ with the desired
electron characteristics.

To guarantee early rejection, the 
HLT Steering can reject an event after any step during the validation 
procedure. If the first step in the example above 
had not resulted in two electron candidates
with the appropriate cluster shapes, the HLT Steering would have stopped the 
validation of the specific $Z$ decay signature under scrutiny without passing 
to step 2. Note also that the HLT processing is not organized
vertically (carry out the full sequence of algorithms to arrive first at a
reconstructed electron for the first RoI and only then do the same for 
the second) but horizontally (carry out only the reconstruction foreseen in 
step x for the first and then the second seed and use the output as seeds for
the next step).

\subsection{Steering}

In this section, the main ingredients of the HLT Steering package,
Trigger Menus, Sequence Tables and Trigger Elements, are discussed briefly. 
For a detailed description, see~\cite{Steering}.

\subsubsection{Trigger Menu and Sequence Table}

\begin{figure*}[t]
\includegraphics[width=135mm]{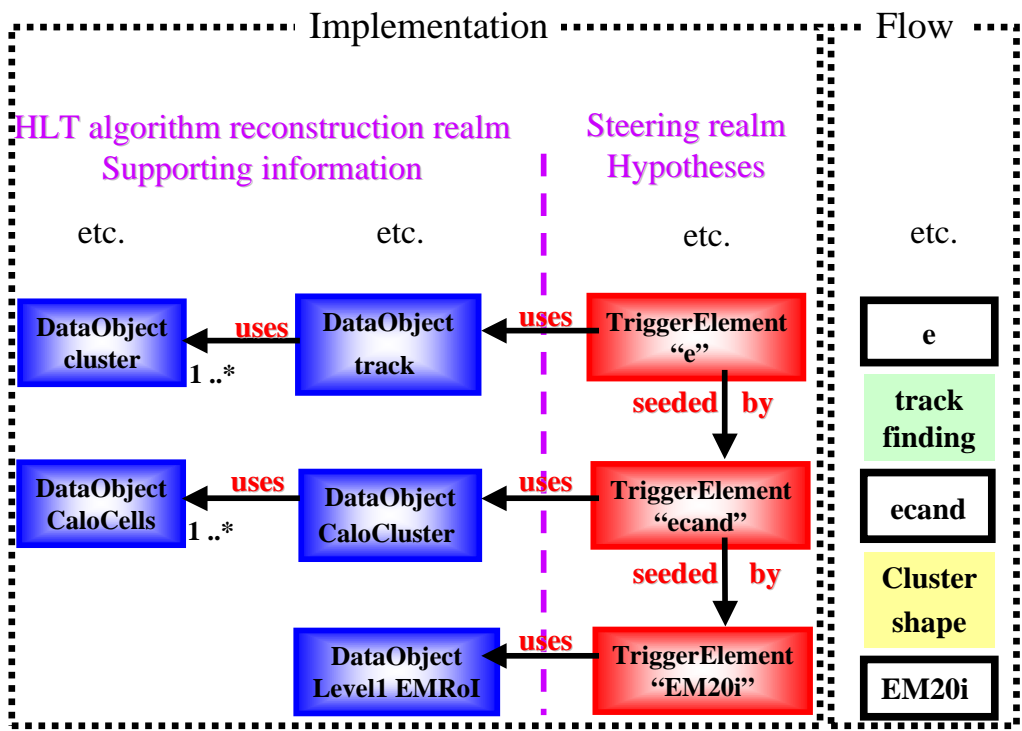}
\caption{Example illustrating the way Trigger Elements are used for 
communication between the HLT Steering and the HLT algorithms. Time increases 
from bottom to top.} 
\label{proc2}
\end{figure*}

In the example above, only one signature (hypothesis) and for this 
signature only one sequence of algorithms and 
intermediate signatures was considered. 
Different from this simple example, the ATLAS HLT accepts an event if it is 
consistent with at least one signature (hypothesis) in a whole catalog of 
possible ones. Also, depending on the Level-1 RoIs, there may be more than one 
possible sequence of algorithms and intermediate signatures per signature.
But as in the example above, the HLT processing is organized horizontally,
not vertically. This means also that the HLT does not examine one signature
and only then the next. Instead, for each signature, the processing is broken
up in single steps and the HLT Steering carries out the processing per step,
not per signature.

Each step is accompanied by one Trigger Menu and one Sequence Table.
A Trigger Menu is a list of all intermediate signatures in its step for which 
an event can be accepted. The Sequence Table specifies which HLT algorithms 
are to be executed, given the seeds as input to the step.

For more details, see~\cite{TriggerMenu}.

The HLT Steering deploys algorithms in a very specific way. 
The task of an HLT algorithm is not to perform a full scan for, e.g., all 
muon candidates in the event, independent of their 
location in the detector. Instead, the algorithm has to carry out a 
localized reconstruction, guided by a seed, limited to the area in the
detector indicated by the seed. Hence, in the same event the HLT Steering
calls the relevant algorithm once per seed, i.e. depending on the number of 
seeds possibly several times.

\subsubsection{Trigger Elements \label{TE}}

Communication between the HLT Steering and the HLT algorithms happens with the
help of Trigger Elements. The HLT Steering hands over the seeds to a
HLT algorithm by means of Trigger Elements. A HLT algorithm returns the 
result of its processing to the HLT Steering in the form of a Trigger Element.

A Trigger Element has a label, but does not hold any data content. It is 
related to Data Objects (RoIs, tracks, clusters, etc.) and other Trigger 
Elements in a navigable way. The signatures in a Trigger Menu consist of 
logical AND/OR combinations of Trigger Elements.

The following example (see Fig.~\ref{proc2}), simplified as compared to the 
example in Fig.~\ref{proc1}, 
illustrates how the HLT Steering uses Trigger Elements
to communicate with HLT algorithms. The Level-1 trigger finds one isolated 
electromagnetic cluster with $p_T > 20$~GeV. Assume for the sake of the 
argument that the ATLAS trigger accepts events with at least one isolated 
electron with  $p_T >$~30~GeV. Then the RoI found by the Level-1 trigger 
is a possible signature for an event of this type.

\begin{figure*}[t]
\includegraphics[width=135mm]{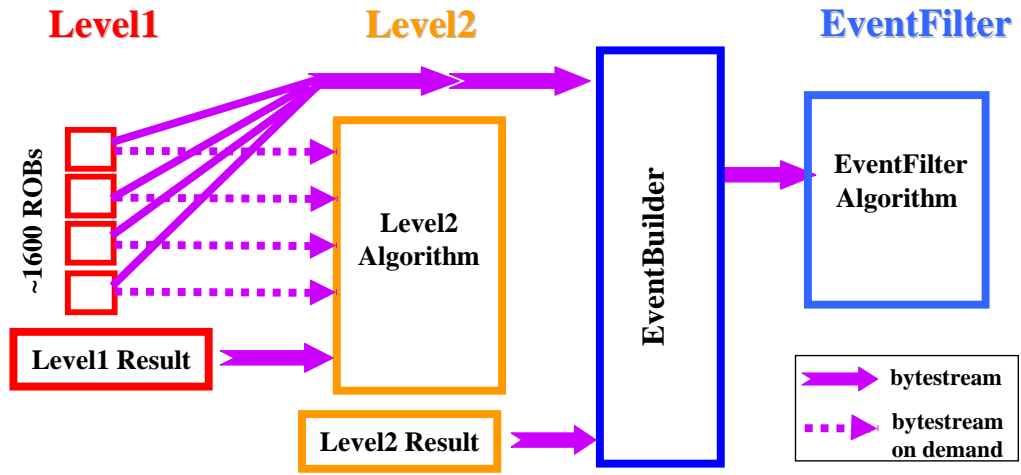}
\caption{Overview of where in the ATLAS trigger raw data are sent to the HLT
in bytestream format.}
\label{proc3}
\end{figure*}

In step 1, the electromagnetic RoI found by the Level-1 trigger is the seed
for the HLT cluster shape algorithm. The HLT Steering creates a Trigger 
Element with label ``EM20i'' that has a navigable ``uses'' relationship to 
the Level-1 RoI as Data Object. The output of the HLT cluster shape algorithm 
is an electron candidate that the algorithm hands back to the HLT Steering as 
a new Trigger  Element with label ``ecand''. This Trigger Element has a 
navigable ``uses'' relationship to the calorimeter cluster that was the result
of the algorithmic processing. Since the HLT cluster shape algorithm was 
seeded by the Level-1 RoI, its corresponding output Trigger Element, ``ecand'',
has a navigable ``seeded by'' relationship with its input Trigger Element,
``EM20i''. Navigable ``uses'' relationships are also possible
between Data Objects. This allows the HLT track finding algorithm in step 3
to access the calorimeter cluster on which its seed is based, and to access 
the calorimeter cells on which the calorimeter cluster is based. 
In addition, via the 
``seeded by'' relationship between its input Trigger Element and the 
input Trigger Element of the preceding step, the algorithm has also access to 
the underlying Data Objects of that previous step.

In the design, the Trigger Elements represent the intermediate hypotheses
that correspond to the intermediate signatures that the HLT Steering examines
to decide whether to continue with the next step of the algorithmic 
processing. The Data Objects, on the other hand, represent the information
supporting these hypotheses.
Differentiating between Trigger Elements and Data Objects has the advantage
of differentiating clearly between the HLT Steering realm,
containing the hypotheses, and the HLT algorithm reconstruction realm,
containing the supporting information.

\subsection{Raw Event Data Model}

Unlike the typical off-line situation where the data are stored in 
a convenient format, e.g. objects in a file or database, 
the HLT event selection software running
on the HLT Linux farms can
access data only in their raw format and needs to invoke the ATLAS Data Flow 
system~\cite{DataFlow1,DataFlow2,DataFlow3,DataFlow4} in order to do so.

The Data Flow system transfers raw data to the HLT in bytestream format.
Figure~\ref{proc3} illustrates where in the ATLAS trigger system raw data is 
exchanged
via bytestreams. In the event of a positive Level-1 trigger decision, the 
Level-2 trigger receives the Level-1 trigger result, in particular the RoIs,
by means of a bytestream. In addition, the Level-2 trigger can demand the 
bytestream fragment from any of the $\sim$1600 ROBs that each hold the raw data of
a part of the ATLAS detector. When the Level-2 trigger accepts an event, the 
Event Builder receives the Level-2 trigger 
result and the bytestream from all $\sim$1600 ROBs and it passes the 
fully-built event by way of another bytestream to the Event Filter.

\subsubsection{Raw Data Objects and Detector Elements}

\begin{figure}[!b]
\includegraphics[width=65mm]{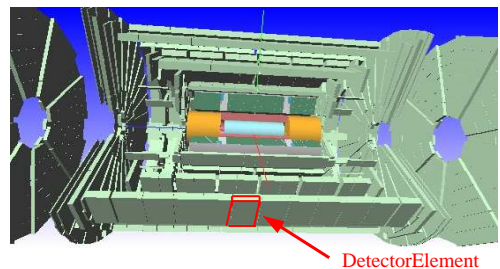}
\caption{Definition of a DetectorElement for the example of the ATLAS Muon 
Spectrometer.} 
\label{proc4}
\end{figure}

In order to allow the HLT algorithms access to the raw data information in 
these bytestreams, this information has to be represented in object form,
as Raw Data Objects (RDOs). The HLT algorithms access the raw data
information by means of the RDOs and the transient event store, where the RDOs
are stored. 

The Raw Event Data Model (Raw EDM) is part of the HLT EDM and specifies the 
design of the RDOs for each
ATLAS subdetector and each bytestream in the Data Flow system. The Raw EDM has
to map onto one another data of very different granularity. The Data Flow 
system provides data with a minimum granularity of one ROB, while a HLT 
algorithm typically needs data organized by geometric detector unit.

For example, in the ATLAS Muon Spectrometer, a geometric detector unit of 
interest for a
HLT selection algorithm is a chamber. A chamber is a physical unit, for
example of Monitored Drift Tubes (MDT), that share the same physical support 
structure, are installed as a whole and may be described by a single
set of condition (e.g. alignment, etc.) constants.
The Raw EDM specifies for all ATLAS subdetectors the object representation of
the relevant geometric detector units, the Detector Elements 
(see Fig.~\ref{proc4}).

In the case of the MDT chambers of the Muon Spectrometer, the Detector 
Element corresponds to a chamber. In the same way as a chamber can be viewed
as a collection of readout channels, a Detector Element object is a collection 
of RDO objects which each hold the information of one readout channel.

\begin{figure}[tb]
\includegraphics[width=65mm]{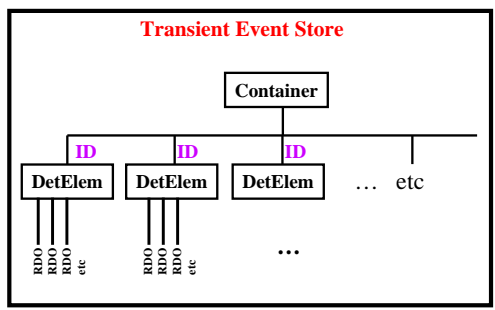}
\caption{Organization of raw data information in the transient event store by 
means of Detector Element objects of which each is a collections of RDO 
objects.} 
\label{proc5}
\end{figure}

RDOs are stored in the transient event store ordered according to 
Detector Elements. The transient store contains for each subdetector a
container of Detector Elements (see Fig.~\ref{proc5}). Each Detector Element 
has a 
unique identifier~\cite{ID} by which it can be retrieved from the store, thus 
allowing 
with one request to the transient store to retrieve collections of RDOs, which 
respresent collections of readout channels in geometric detector units.

\subsection{Bytestream converters \label{BSC}}

\begin{figure*}[tb]
\includegraphics[width=135mm]{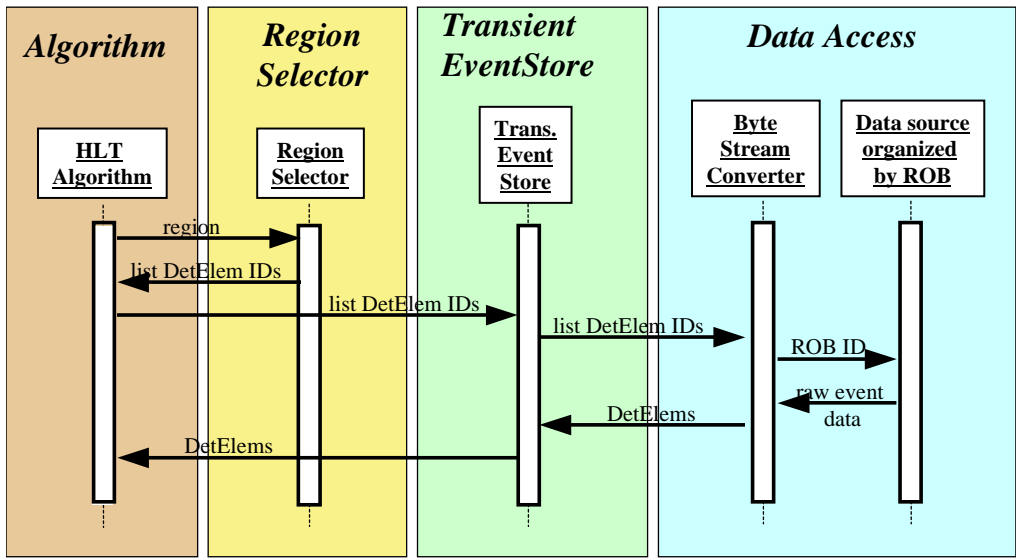}
\caption{HLT algorithm access to raw data information by means of Bytestream 
Converters and the transient event store.} 
\label{proc6}
\end{figure*}

The raw data in the bytestream format in which they are held by the ROBs
can be viewed as a representation of the raw data information in an
event. The RDOs, organized by Detector Element, are another representation
of this information. The conversion from bytestream representation to RDO
representation is the task of Bytestream Converters.
They carry out the conversion on demand of an HLT algorithm.
Their implementation consists of a general infrastructure part that uses
subdetector specific methods for decoding the information from bytestream
format to RDO format. The general part is, for example, responsible for 
storing the RDOs in the correct Detector Element order in the transient store.

\begin{figure}[tb]
\includegraphics[width=65mm]{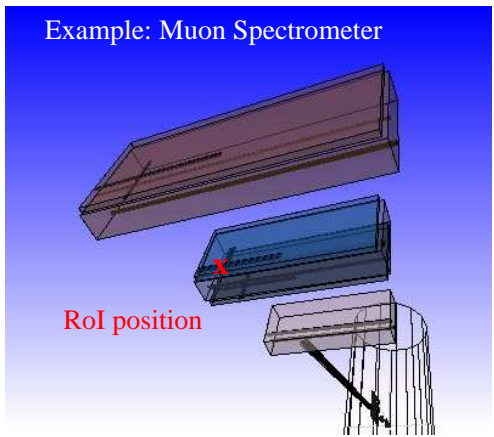}
\caption{Example illustrating the task of the Region Selector.} 
\label{proc7}
\end{figure}

Raw data access via Bytestream Converters proceeds in the following way 
(see Fig.~\ref{proc6}).
The HLT Steering calls an HLT algorithm by passing to it a Trigger Element
with a navigable link to a RoI Data Object that contains the RoI position.
This RoI can, for example, be located in the Muon Spectro\-meter.
The HLT algorithm typically needs access only to the raw data in those 
Muon Spectrometer chambers within a certain region around the
RoI position (see Fig.~\ref{proc7}). These chambers correspond to a certain 
set of Detector Elements.
The task of identifying this set of Detector Elements, given a region in the 
pseudorapidity $\eta$ and the polar angle $\phi$, is performed by the Region 
Selector~\cite{RegionSelector}. The Region Selector returns to the 
calling HLT algorithm the
list of identifiers of the appropriate set of Detector Elements. The HLT
algorithm requests the Detector Elements from the transient store with the help
of this identifier list.

At this stage, two situations can occur. If the requested Detector Elements
are already available in the transient store, the store simply returns them.
If they are not available, however, the corresponding raw data information 
first has to be requested from the Data Flow system. In this case, the 
transient event store invokes the correct Bytestream Converter, which in turn 
requests the raw data in bytestream format from the appropriate ROBs. 
From the succession of 32-bit words sent by each ROB, the Bytestream Converter 
extracts the relevant information and uses it to initialize the collections of 
RDOs. The thus obtained collections of RDOs, or Detetcor Elements, are stored
in the transient store, which passes them on to the HLT algorithm. 

The 
extraction of raw data information into RDOs can be supplemented by
additional data preparation algorithms, e.g. cluster finders for the 
Silicon Trackers. The object representation of extracted and 
prepared raw data are called Reconstruction Input Objects. Their definition is 
part of the HLT EDM. They are 
organized inside the transient store in collections as are RDOs, and can be 
requested by an HLT algorithm in the same way as RDOs.

\section{OFF-LINE CODE IN ON-LINE CONTEXT}

A high level of integration of the code designed for off-line use and the code 
designed for on-line use is clearly a desirable design goal. Ideally, 
it should be possible to develop and maintain on-line software on simulated and 
real data in a well-tested, flexible and user-friendly environment as is 
offered by the off-line software. This does not only apply to the HLT 
algorithms, but also to the HLT infrastructure code used by them.
Furthermore, it clearly enhances the HLT flexibility in dealing with changes 
in the LHC running conditions when it is possible to migrate software
freely between off-line environment and Event Filter and between Event Filter 
and Level-2 trigger.

In the previous sections, the common software infrastructure of Level-2 
trigger software and Event Filter software has been described. 
In order to use this HLT software infrastructure and the HLT algorithms 
in an off-line environment, the HLT software has to 
comply with the basic design principles of the ATLAS off-line framework.

The ATLAS off-line framework, Athena~\cite{Athena}, is based on a 
GAUDI~\cite{GAUDI} (developed by the LHCb
experiment) core. As a consequence, Athena has adopted a number of the
basic principles of the GAUDI design. Examples are the separation of the data 
from the algorithms producing and consuming the data, separation of the 
transient and the persistent representation of data and the use of converters 
that convert one data representation into another.

For the HLT event selection software this means, for example, 
that HLT algorithms are forced to
communicate indirectly, via the transient event store. 
This requirement is met by the use of Trigger Elements, as described in 
Sec.~\ref{TE}. It also means that 
persistent data can be made available to the HLT event selection software only
by means of converters that convert them into a transient representation that 
is then stored in the transient event store, from where HLT algorithms can 
request data. The Bytestream Converters discussed in Sec.~\ref{BSC} and their 
use as illustrated in Fig.~\ref{proc6} comply with this requirement. 
Furthermore, using the HLT event selection software in Athena
implies at the very least common interfaces to general services, and may even 
mean using the very same services, e.g. database access tools, geometry service
etc., by Athena and by the on-line framework. 

The HLT event selection software also needs to comply with the much more
stringent on-line performance requirements concerning speed and robustness.
Hence, also any off-line-imposed elements or code elements imported from the
off-line environment must meet the on-line
performance requirements. They are considerably more stringent than would be
otherwise needed for pure off-line use.

The following section describes the current off-line dependencies 
introduced into the HLT software for the above mentioned reasons.
The experience gained so far in the attempt of using off-line code in the 
on-line context of the trigger is discussed.

\begin{figure*}[tb]
\includegraphics[width=135mm]{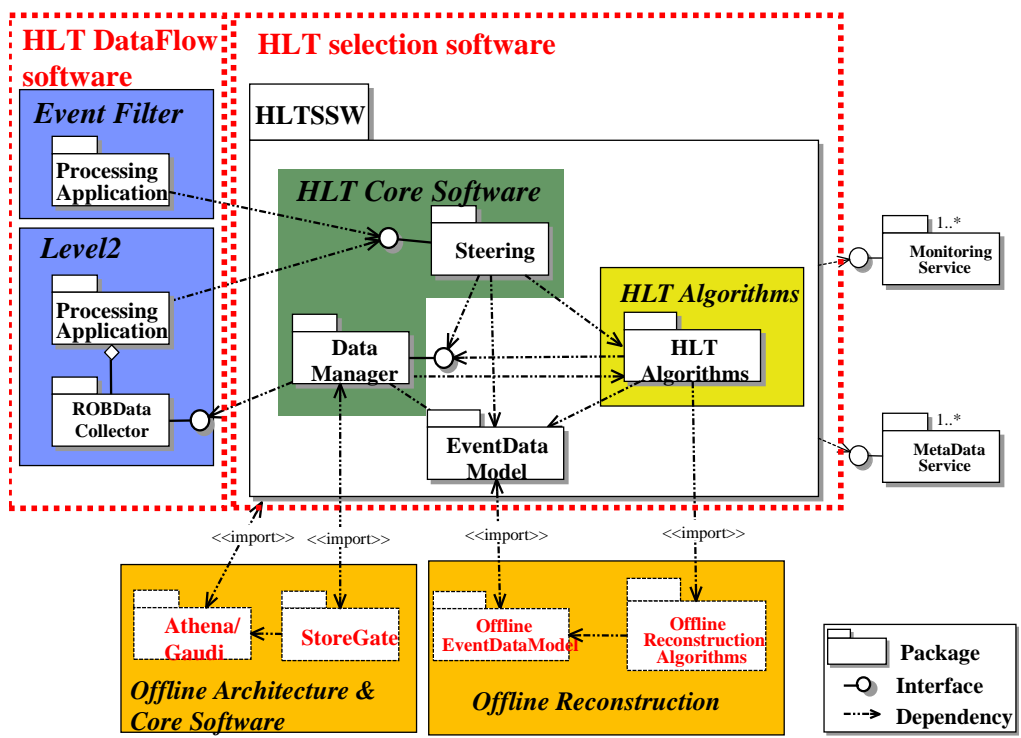}
\caption{The four components of the HLT selection software and their 
dependencies on off-line software.} 
\label{proc8}
\end{figure*}

\subsection{Current Off-Line Dependencies} 

A high level of integration between off-line and on-line code can be achieved in
different ways. One possibility consists in choosing the interfaces in the 
off-line and in the on-line code such that on-line software can be used in the 
off-line framework and vice versa. Another possibility is to use the same
framework for off-line and on-line purposes. In the concrete ATLAS situation, 
this means using Athena as on-line framework as well. In addition, any 
combination of these two extremes is possible.

In the current implementation of the HLT software, Athena is used as framework
in the Event Filter. Level-2 employs a specialized framework that uses
interfaces compatible with those in Athena, so that any Level-2 infrastructure 
and algorithm code can also be used within Athena (see next section). 
The HLT infrastructure code
(e.g. transient event store, Bytestream Converters, etc.) currently re-uses 
code developed in Athena.
  
Figure~\ref{proc8} depicts the current off-line dependencies in the HLT event 
selection software. The HLT framework depends on the off-line framework.
For the Event Filter this means re-use of Athena as framework, for the Level-2
trigger this means use of a slightly modified version (see next section) 
of the Athena core software, based on GAUDI. The Data Manager in the current 
implementation is the off-line transient event store, 
StoreGate~\cite{StoreGate}. The HLT EDM
re-uses the off-line EDM. In the Event Filter, off-line algorithms are re-used
as selection algorithms. In the Level-2 trigger, specialized Level-2 algorithms
are employed which, however, are set up such that they can also be run in the 
Athena framework.

The execution of the HLT event selection software is controlled by a 
Processing Application that is part of the HLT Data Flow and Data Acquisition 
software.

\subsection{The Special Level-2 Environment}

The performance constraints imposed by the HLT are most stringent for the 
Level-2 trigger. There, the average latency available to the HLT selection 
software for accepting or rejecting an event is $\sim$10~ms. The Level-2 software
needs to be able to handle a data input rate of up to 75~kHz and to sustain 
a data output rate of $\mathcal{O}(2)$~kHz error-free and deadtime-free for
extended periods of time.
Requirements of such stringency with respect to speed and robustness are
usually not met by software designed for pure off-line use.
An additional complication arises from the need for multi-threading in the 
Level-2 trigger.

\begin{figure}[tb]
\includegraphics[width=65mm]{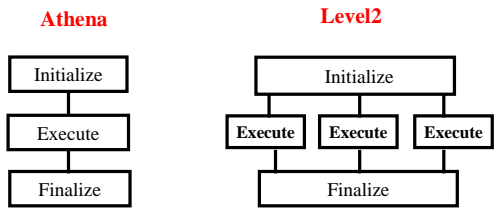}
\caption{Difference between the off-line framework Athena and 
the framwork needed for the Level-2 trigger, where multi-threading is
required.
} 
\label{proc9}
\end{figure}

In order to keep the idle time of the CPUs in the Level-2 Linux PC farm
to a minimum, each CPU processes in parallel three different events, i.e.
each Level-2 CPU carries out at the same time three execute loops.
This is in marked difference to the situation foreseen in Athena, where 
the initialization phase is followed by a single execute loop that ends in a 
finalization phase (see Fig.~\ref{proc9}). 
In the case of running inside Athena,
Athena controls the event execute loop; in the multi-threaded 
Level-2 software the control needs to be with the HLT Data Flow software.
In addition, in the Level-2 software, all algorithms and services 
have to be configured and initialized strictly outside of the event execute 
loop. Hence Athena cannot be used as HLT framework
for the Level-2 trigger.  

In order to comply nonetheless with the design goal that the Level-2
event selection software be usable with the Athena framework for
development and maintenance purposes, an interface layer between the 
HLT selection and the HLT Data Flow software was implemented, the 
so-called Steering Controller~\cite{GAUDIL2}. The Steering Controller uses only a minimal 
set of GAUDI features. In addition, it uses a modification to the GAUDI base 
libraries that allows thread-specific instantiation. The standard GAUDI 
command $EventLoopMgr \rightarrow executeEvent()$ is, for example, replaced by
$EventLoopMgr \_ (threadID) \rightarrow executeEvent()$, where $threadID$ 
corresponds to the number of the thread (1, 2, 3). Accordingly, GAUDI
services are called with the command $AnyServiceNeededByAlgo \_ (threadID)$ and
algorithms with $AnyHLTALgorithm \_ (threadID)$, etc. The possibility of
thread-specific instantiation has since been included in GAUDI as a standard
feature. The Steering Controller can be used within the off-line framework 
Athena because in a single-thread environment the $\_(threadID)$ suffix 
collapses
to a blank. Beyond this the Steering Controller uses only standard GAUDI
features, which are all contained in the GAUDI-based Athena.

The special Level-2 feature of running in multi-thread mode may cause problems 
when using standard software tools as, e.g., the Standard Template Library 
(STL), even when the software components used are in principle thread-safe. An 
unexpected problem of this kind arose with the specific implementation 
of STL containers in use in the HLT code. The container implementation, though
thread-safe, proved highly thread-inefficient by setting unnecessary locks.
The problem was overcome by modifying by hand the container allocator 
implementation~\cite{MultiThreading}.

It should be noted that the problems arising from the use of multiple threads
in the Level-2 trigger software are not present for the Event Filter. 
The Event Filter uses multiple threads, but the Data Flow and HLT event 
selection software are set up in a different way.
As a result, it is possible to use Athena as is as the framework for running 
HLT selection algorithms in the Event Filter~\cite{AthenaEF}. However, 
when Athena was 
used for the first time in the Event Filter, it became aparent that an 
inacceptable 30\% of the total processing time per event was spent in the 
off-line Raw EDM available at that time. This problem has since been remedied,
but illustrates the need for a thorough evaluation of the suitability of
off-line code for use in the HLT software. It also illustrates the need for a 
close and continuous collaboration between the HLT (``on-line'') and the 
``off-line'' communities in order to adopt successfully ``off-line'' software
for ``on-line'' use.

\section{CONCLUSIONS AND OUTLOOK} 

The architecture of the HLT selection software described in this article
is currently being validated by implementing the software for full vertical 
trigger slices. Two vertical slices
are currently under development, an electron/gamma slice and a muon slice.
The vertical trigger slices comprise the software to simulate the full chain 
of trigger decisions, starting with the Level-1 trigger, that leads to 
identifying an event as containing a single electron, gamma or muon.
Integration of the HLT selection software with the other components of the HLT 
software is on-going with the goal of running the electron/gamma and muon
vertical slices in testbeds of the foreseen Linux PC farms. Detailed 
measurements of timing and physics performance of each part of the 
HLT software are under way and will be reported in the HLT Technical Design 
Report in summer 2003.

A central goal of the current implementation of the HLT selection software
is a high level of integration between off-line and on-line code.
This goal has sparked off a very fruitful collaboration between the
HLT and the on-line software developer communities.
The attempt of adapting and re-using off-line code in the HLT software has 
been carried furthest in the specific way of accessing raw data in the HLT by
means of Bytestream Converters that are hidden behind a call to the off-line 
transient event store, StoreGate.

Given the ever increasing available CPU speed, allowing an ever higher level
of abstraction, the concept of re-using off-line code in the on-line context 
appears a logical development. However, given the stringent constraints
and performance requirements of the HLT, the present implementation of the
HLT selection software and the heavy re-use it makes of off-line software
is clearly experimental. This is in particular true for the Level-2 trigger.
The results of the on-going validation effort will show if the chosen 
ansatz is sustainable.

\section{ACKNOWLEDGMENT} \label{Ack}
We would like to acknowledge the help and support of the ATLAS
Data Acquisition Group and the ATLAS Off-Line and Detector software groups.

\end{document}